\documentclass{aa}

\usepackage{graphicx, psfig, times, amsfonts}

\newcommand{\less}{\raisebox{-1.1mm}{$\stackrel{<}{\sim}$}}
\newcommand{\more}{\raisebox{-1.1mm}{$\stackrel{>}{\sim}$}}

\newcommand{\M}{{\sc 2mass}}
\newcommand{\OG}{{\sc ogle}}
\newcommand{\DE}{{\sc denis}}

\begin{document}

\title{Eclipsing binaries in the Galactic Bulge: \\ candidates for distance estimates
\thanks{Table~2 is available in electronic form at the CDS via
anonymous ftp to cdsarc.u-strasbg.fr (130.79.128.5) or via
http://cdsweb.u-strasbg.fr/cgi-bin/qcat?J/A+A/. 
Figure~1 and Appendices~A and B are available in the on-line edition of A\&A.}  
}

\author{
M.A.T. Groenewegen
}

\institute{
Instituut voor Sterrenkunde, K.U. Leuven, PACS-ICC, Celestijnenlaan 200B, 
B--3001 Leuven, Belgium 
}

\date{received: 2005,  accepted: May 4th, 2005}

\offprints{Martin Groenewegen (groen@ster.kuleuven.ac.be)}


\abstract{
The 222~000 $I$-band light curves of variable stars detected by the
\OG\--{\sc ii} survey in the direction of the Galactic Bulge have been
searched for eclipsing binaries (EBs). A previously developed code to
analyze lightcurve shapes and identify long period variables (LPVs) has been
adapted to identify EBs. The parameters in the modified code have been
optimised to recover a list of about 140 detached EBs in the Small
Magellanic Cloud previously identified in the literature as particularly 
well suited for distance estimates (and wich have periods \more 0.85 days). 
The power of the code is demonstrated by identifying 16 and
178 previously uncatalogued EBs in the SMC and LMC, respectively.
Among the 222~000 variable stars in the direction of the Galactic
Bulge 3053 EBs have been identified. Periods and phased
lightcurves are presented.
\keywords{Stars: distances - binaries: eclipsing - Galaxy: bulge - distance scale}
}


\maketitle

\section{Introduction}

Detached double-lined eclipsing binaries (hereafter EBs for short) are
very suitable as primary distance indicators when accurate photometry
and radial velocity (RV) data are combined (e.g. Andersen
1991). The massive photometric monitoring data obtained in the course
of the micro lensing surveys in the 1990's of the Small and Large
Magellanic Clouds has revealed many candidate detached eclipsing
binaries based on the shape of the lightcurves (Grison et al. 1995,
Alcock et al. 1997, Udalski et al. 1998, Bayne et al. 2002,
Wyrzykowski et al. 2003, 2004).

Some of them have been followed-up with spectroscopy to obtain RV data
and perform a combined lightcurve and RV curve analysis to obtain the
distance to the system (see e.g., Guinan et al. 1998, Hilditch et
al. 2005, and the review by Clausen 2004 for a list of systems with
recent distance estimates).  What makes EBs particularly powerful is
that the error in the distance estimate in a single well-observed
system (0.06 - 0.10 in distance modulus, see Clausen 2004) is already
comparable to the dispersion in optical and infrared Cepheid
$PL$-relation based on hundreds of stars (e.g. Sandage et al. 2004,
Nikolaev et al. 2003).

In the present paper a list of eclipsing binaries in the direction of
the Galactic Bulge (GB) is presented based on analysis of \OG\--{\sc ii} data. 
Previously, 1650 EBs have been identified in \OG\--{\sc i} data
(Udalski et al. 1994, 1995a, 1995b, 1996, 1997).

The paper is structured as follows. In Section~2 the \OG\--{\sc ii}
surveys is briefly described. In Section~3 the model to identify EBs
from the lightcurve shape is briefly presented. The results are presented in
Section~4, and discussed in Section~5.

\section{The data sets}

The \OG\--{\sc ii} micro lensing experiment observed fourty-nine fields
in the direction of the GB. Each field has a size 14.2\arcmin$\times$57\arcmin\ 
and was observed in $BVI$, with an absolute photometric accuracy of
0.01-0.02 mag (Udalski et al. 2002).  Table~\ref{tab-ogle} lists the
galactic coordinates of the field centers and the total number of
sources detected in these fields.

Wozniak et al. (2002) present a catalog of about 222~000 variable
objects based on the \OG\ observations covering 1997-1999, applying
the difference image analysis (DIA) technique on the $I$-band data.
The data files containing the $I$-band data of the candidate variable
stars was downloaded from the \OG\ homepage
(http://sirius.astrouw.edu.pl/$^{\sim}$ogle/).

\begin{table*}
\caption{Properties of the \OG-fields and the number of detected EBs}


\begin{tabular}{rrrrrlr} \hline
BUL\_SC & $l$ & $b$ & Total$^{(a)}$ & Variable$^{(b)}$ & $A_{\rm V}$$^{(c)}$ 
                                          &  EBs$^{(d)}$  \\ 
        &   &   &    &      &             &           \\ \hline
1  &  1.08 & -3.62 & 730   & 4597 & 1.68 / 1.49 &  77 \\
2  &  2.23 & -3.46 & 803   & 5279 & 1.55 / 1.65 &  70 \\
3  &  0.11 & -1.93 & 806   & 8393 & 2.89        & 130 \\
4  &  0.43 & -2.01 & 774   & 9096 & 2.59 / 2.94 &  146 \\
5  & -0.23 & -1.33 & 434   & 7257 & 5.73 / -- / 4.13       & 39 \\
6  & -0.25 & -5.70 & 514   & 3211 & 1.37        &  41 \\
7  & -0.14 & -5.91 & 463   & 1618 & 1.33 / 1.28 &  21 \\
8  & 10.48 & -3.78 & 402   & 2331 & 2.14        &  29 \\
9  & 10.59 & -3.98 & 330   & 1847 & 2.08        &  18 \\
10 &  9.64 & -3.44 & 458   & 2499 & 2.23        &  36 \\
11 &  9.74 & -3.64 & 426   & 2256 & 2.27        &  25 \\
12 &  7.80 & -3.37 & 535   & 3476 & 2.29 / 2.20 &  36 \\
13 &  7.91 & -3.58 & 570   & 3084 & 2.06 / 1.82 &  41 \\
14 &  5.23 &  2.81 & 619   & 4051 & 2.49        &  83 \\
15 &  5.38 &  2.63 & 601   & 3853 & 2.77        &  41 \\
16 &  5.10 & -3.29 & 700   & 4802 & 2.15 / 2.23 &  48 \\
17 &  5.28 & -3.45 & 687   & 4690 & 1.94 / 2.29 &  46 \\
18 &  3.97 & -3.14 & 749   & 5805 & 1.83        &  78 \\
19 &  4.08 & -3.35 & 732   & 5255 & 2.74        &  61 \\
20 &  1.68 & -2.47 & 785   & 5910 & 1.94 / 2.02 & 119 \\
21 &  1.80 & -2.66 & 883   & 7449 & 1.83 / 1.78 & 116 \\
22 & -0.26 & -2.95 & 715   & 5589 & 2.74        &  91 \\
23 & -0.50 & -3.36 & 723   & 4815 & 2.70        &  68 \\
24 & -2.44 & -3.36 & 612   & 4304 & 2.52        &  68 \\
25 & -2.32 & -3.56 & 622   & 3046 & 2.34        &  77 \\
26 & -4.90 & -3.37 & 728   & 4713 & 1.86        &  76 \\
27 & -4.92 & -3.65 & 691   & 3691 & 1.69        &  66 \\
28 & -6.76 & -4.43 & 406   & 1472 & 1.64        &  26 \\
29 & -6.64 & -4.62 & 492   & 2398 & 1.53        &  44 \\
30 &  1.94 & -2.84 & 762   & 6893 & 1.91 / 1.78 & 137 \\
31 &  2.23 & -2.94 & 790   & 4789 & 1.81 / 1.74 &  87 \\
32 &  2.34 & -3.14 & 797   & 5007 & 1.61 / 1.82 &  99 \\
33 &  2.35 & -3.66 & 739   & 4590 & 1.70 / 1.82 &  44 \\
34 &  1.35 & -2.40 & 961   & 7953 & 2.52 / 2.32 & 127 \\
35 &  3.05 & -3.00 & 771   & 5169 & 1.84 / 2.20 &  64 \\
36 &  3.16 & -3.20 & 873   & 8805 & 1.62 / 1.52 &  85 \\
37 &  0.00 & -1.74 & 664   & 8367 & 3.77        & 119 \\
38 &  0.97 & -3.42 & 710   & 5072 & 1.83 / 1.94 & 119 \\
39 &  0.53 & -2.21 & 784   & 7338 & 2.63 / 2.70 & 144 \\
40 & -2.99 & -3.14 & 631   & 4079 & 2.94        &  48 \\
41 & -2.78 & -3.27 & 603   & 4035 & 2.65        &  55 \\
42 &  4.48 & -3.38 & 601   & 4360 & 2.29        &  67 \\
43 &  0.37 &  2.95 & 474   & 3351 & 3.67        &  34 \\
44 & -0.43 & -1.19 & 319   & 7836 & 6.0 / -- / 6.00        & 32 \\
45 &  0.98 & -3.94 & 627   & 2262 & 1.64 / 1.53 &   4 \\
46 &  1.09 & -4.14 & 552   & 2057 & 1.71 / 1.65 &   3 \\
47 & -11.19 & -2.60 & 301  & 1152 & 2.60        &   7 \\
48 & -11.07 & -2.78 & 287  &  973 & 2.35        &   9 \\
49 & -11.36 & -3.25 & 251  &  826 & 2.09        &   6 \\
total &     &     & 30490  & 221701     &       & 3053      \\ 
\hline
\end{tabular}
\label{tab-ogle}

(a) Total number of objects detected in the field. From Udalski et
al. (2002), in units of $10^3$ objects

(b) Total number of candidate variable stars. From Wozniak et al. (2002).

(c) Visual extinction. From Sumi (2004), except for SC44,
where $A_V$ = 6.0 has been adopted based on the proximity to SC5.
The second value--when listed--comes from Popowski et al. (2003).
The third value--when listed--comes from Schultheis et al. (1999).

(d) Total number of EBs.

\end{table*}

\section{ Analysis of the lightcurve shape and selecting eclipsing binaries}

The model to analyse the lightcurve shape and identify long period
variables (LPVs) is described in detail in Appendices~A-C in
Groenewegen (2004; herafter G04).

Briefly, a first code (see for details Appendix~A in G04) sequentially
reads in the $I$-band data for the objects, determines periods through
Fourier analysis, fits sine and cosine functions to the light curve
through linear least-squares fitting and makes the final correlation
with the pre-prepared \DE\ and \M\ source lists. All the relevant
output quantities are written to file. This part of the code is
adapted significantly to better deal with the specific properties of
eclipsing binaries w.r.t. pulsational variables, as described in
Appendix~A of the present paper. Brief, phase dispersion minimization
is used next to Fourier analysis to select the true orbital period,
and the EBs candidates are selected based on statistical properties of
the phased light curve.

The output file of the first code is read in by the second code (see
for details Appendix~B in G04).  A further selection may be applied
(typically on period, amplitude and mean $I$-magnitude), multiple
entries are filtered out (i.e. objects that appear in different \OG\
fields), and a correlation is made with pre-prepared lists of objects
for cross-identification. The output of the second code is a list with
EB candidates.

The third step (for details see Appendix~C in G04) consists of a
visual inspection of the fits to the (phased) light curves of the
candidate EBs and a literature study through a correlation with the
{\sc simbad} database. Non-EBs are removed, and sometimes the fitting
is redone. The final list of EB candidates is compiled.

\section{Results}

\subsection{Testing the code on selected SMC data}

In order to set and finetune the parameters described in Appendix~A
that define the selection of an EB, and also to demonstrate that with
these particular settings EBs can indeed be retrieved, the code is
tested and run on the \OG\ data of variable stars in the Magellanic
Clouds (hereafter MCs; Zebru\'n et al. 2001).

As the ultimate aim of the project is to define EBs in the GB suitable
for distance determinations the parameters that define the selection
of EBs (3 statistical parameters that define the shape of the phased
lightcurve, and the parameters {\it significance} and {\it hifac} of
the {\sc numerical recipes} (Press et al. 1992) subroutine {\it fasper}) 
are tuned to retrieve the stars in the lists of ``ideal'' distance
indicators in the SMC previously selected by Udalski et al. (1998) and
Wyithe \& Wilson (2001) based on the first release of \OG\ data (based
on a shorter timespan of \OG\ data and aperture photometry instead of DIA). 
What constitutes an ``ideal'' EB in terms of a distance
indicator is difficult to quantify exactly. Typically the system
should be well-detached with small values of the stellar radii
relative to the semi-major axis ($r/a \less 0.2)$ and have deep
eclipses (e.g. Wyithe \& Wilson, 2001). 

Not all of the stars from  Udalski et al. (1998) and
Wyithe \& Wilson (2001)  are listed in the Zebru\'n et al. (2001) data set. 
In fact, Wyrzykowski et al. (2004) state that of the 1527 EBs listed
by Udalski et al. (1998) only 935 can be found in the DIA catalog of
variable stars by Zebru\'n et al. (2001), and quote the incompleteness
of the DIA catalog for faint stars as the probable reason. Of the 153
stars listed by Udalksi et al. (1998) and the 22 unique additional
objects in Tables~4 and 5 of Wyithe \& Wilson (i.e. not already listed
in Udalski et al.), 142 are found back by Zebru\'n et al. (2001), and
these stars represent the test data set.

Of these 142, 137 are correctly identified as EBs by the program using
the parameters mentioned in Appendix~A. One object (OGLE005102.88-730941.0) 
is found with the period as listed by Udalski et al. (1998) but the
phased light curve is poor with the DIA data set, and the object is
not suitable as a distance indicator from the present data. Two stars
would have been correctly identified as EBs were it not that the
significance of the peak of the Fourier spectrum is above the adopted
threshold. The other two are missed because the parameters that are
being used to describe the shape of the phased lightcurve are outliers. 
At the end of the day, it is deemed acceptable to miss 4 stars out of
a 141 suitable candidates. It should be mentioned that these ``ideal''
distance indicators have orbital periods larger than 0.85 days and so
the code is optimised to be sensitive to this period range, and hence
will be biased against shorter orbital periods.

\subsection{Applying the code to SMC and LMC data}

As a further test, the code is applied to the full set of 68~000
lightcurves of variable stars in the SMC and LMC by Zebru\'n et
al. (2001), and compared to the outcome of the previous work by
Wyrzykowski et al. (2003, 2004), based on the same dataset as is the
present work.

Wyrzykowski et al. (2003) find 2580 EBs in the LMC, and Wyrzykowski et
al. (2004) find 1351 EBs in the SMC, based on an artificial neural
network.  Note that they attempted to find all types of eclipsing
binaries, while the aim of the present paper is to detect EBs
potentially suitable for distance determination, hence preferentially
detached systems.

Applying the code resulted in 1856 LMC and 752 SMC candidate EBs.

For the SMC, 714 of the 752 objects are listed in Wyrzykowski et
al. (2004) as EBs. Of the 38 not listed in Wyrzykowski et al., 20 are
eliminated at the visual inspection stage as the phased lightcurves
are not typical of that of an EB; 13 are in fact known Cepheids. Of
the 18 remaining stars, 2 are classified as EB by MOA (Bayne et al. 2002),
and 16 are new EBs not previously identified. Names, periods and the
phased lightcurve are presented in Appendix~B.

For the LMC, 1616 of the 1856 objects are listed in Wyrzykowski et
al. (2003) as EBs. Of the 240 not listed in Wyrzykowski et al., 51 are
eliminated at the visual inspection stage as the phased lightcurves
are not typical of that of an EB; 7 are known Cepheids, 2 are known RR
Lyrae and 24 are LPVs. Of the 189 remaining stars, 11 are classified as EB
by MACHO (Alcock et al. 1997), and 178 are new EBs not previously identified. 
Names, periods and the phased lightcurve are presented in Appendix~B.

The conclusion is that the simple method developed for detecting EBs is
complementary to the neural network method used in Wyrzykowski et
al. (2003, 2004) as, considering both MCs, about 7.5\% of the detected
objects are newly discovered EBs. The number of false candidates
(mostly cepheids, RR Lyrae and LPVs) is about 2.5\%.

\subsection{Applying the code to the Galactic Bulge}

The numerical code is applied to the \OG\--{\sc ii} data in the direction of the Galactic Bulge.
After visual inspection of the lightcurves a sample of 3053
objects remain. The largest number of objects that are removed at
this stage are LPVs especially at longer periods ($P_{\rm binary}$
\more 150, respectively $P_{\rm pulsation}$ \more 75 days) and EBs
with orbital periods \less 0.85 days where an alias frequency is picked
up, which nevertheless phases well enough to fulfill the initial selection criteria.

\begin{figure*}

\includegraphics{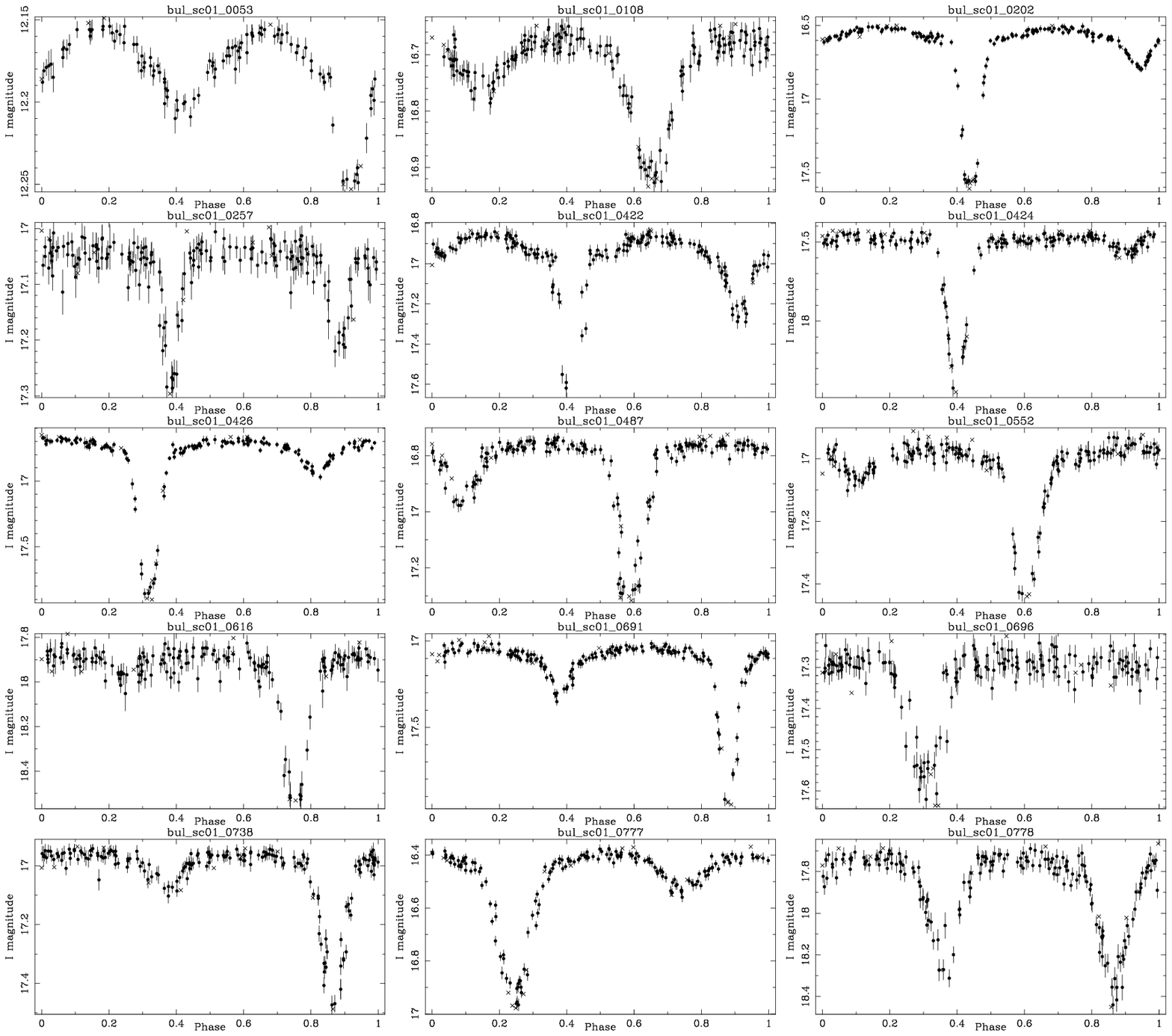}
\caption[]{
First 15 entries of the electronically available figure with all
phased lightcurves of EBs in the direction of the Galactic Bulge. 
Crosses indicate data points not included in the analysis. Phase zero
is given by the Julian Date listed in Table~\ref{TAB-A}.
}
\label{Fig-A}
\end{figure*}

The number of objects per field is listed in the last column of
Table~\ref{tab-ogle}. Table~\ref{TAB-A} lists the 3053 objects
with periods, and Figure~\ref{Fig-A} displays the phased lightcurves. The
typical error in the period is 3 $\times 10^{-4} P$. Only one object
is listed in the SIMBAD database, namely bul\_sc26\_3510 (R.A.=
$17h47m28.25s$, Dec= $-34d46m31.0s$), a.k.a. MM5-A V47 in Udalski et
al. (1997) and classified as a W UMa-type binary. The period quoted
there is 1.10577$d$ while in the present analysis 1.105801$d$ is derived.

%

\begin{table}
\caption[]{
First entries in the electronically available table, which lists:
OGLE-name, oribital period, and time used to define phase zero.
}

\begin{tabular}{crr} \hline
         OGLE name    & Period ($d$) & $T^{(a)}$ \\ \hline

                      bul\_sc01\_0053 &   2.521648 & 530.819 \\
                      bul\_sc01\_0108 &   1.532373 & 550.789 \\
                      bul\_sc01\_0202 &   4.513265 & 535.895 \\
                      bul\_sc01\_0257 &   2.378940 & 540.765 \\
                      bul\_sc01\_0422 &   4.201144 & 530.836 \\
                      bul\_sc01\_0424 &   1.738786 & 530.836 \\
                      bul\_sc01\_0426 &   4.236257 & 530.819 \\
                      bul\_sc01\_0487 &   1.782421 & 550.789 \\
                      bul\_sc01\_0552 &   2.049995 & 550.789 \\
                      bul\_sc01\_0616 &   1.761581 & 530.837 \\
                      bul\_sc01\_0691 &   3.350188 & 535.895 \\
                      bul\_sc01\_0696 &   1.329869 & 550.789 \\
                      bul\_sc01\_0738 &   1.427248 & 540.765 \\
                      bul\_sc01\_0777 &   1.578524 & 550.789 \\
                      bul\_sc01\_0778 &   1.107325 & 540.765 \\
                      bul\_sc01\_0851 &   8.904982 & 550.789 \\

\hline
\end{tabular}

(a) $T$ = (JD-2450000) used to define zero phase in Figure~\ref{Fig-A}.

First entries only. Complete table available in electronic form at the CDS.

\label{TAB-A}
\end{table}


\section{Summary}

The paper presents a simple method to detect eclipsing binaries based
on the shape of the phased lightcurve. In view of the ultimate aim to
use EBs to derive the distance to the Galactic Centre, the parameters
of the model are finetuned to retrieve a large sample of EBs in the
SMC previously classified in the literature as potentially suitable for distance
determinations. The code is run on the 68~000 \OG\ variable stars in
LMC and SMC to find 194 apparently previously unclassified EBs,
and on the about 222~000 \OG\ variable stars in the direction of the
Bulge to find about 3000 EBs. It should be stressed that due to the
finetuning of the model no short period EBs ($P \less 0.85d$) are among these.

It should also be stressed that altough the code was tuned to
retrieve EBs peviously classified in the literature as suitable for
distance determinations, the converse is not necesarily true: The
objects returned by the code and listed in Tables~\ref{TAB-A} and
\ref{TAB-NewMCs} are not {\em automatically} suitable as potential
distance indicators. For example, none of the 15 objects displayed in
Figure~\ref{Fig-EB-NEW} seems suitable for this purpose. In this case
this is not surprising as only newly discovered EBs, not in the
exhaustive lists of Wyrzykowski et al. (2003, 2004), are listed and so
these objects have relatively low S/N lightcurves. Of the lightcurves displayed 
in Figure~\ref{Fig-A} only a few appear suitable for follow-up studies.

In the future it is planned to select a small sub-sample among
the 3000 objects based on a proper analysis of the lightcurve in
terms of radii determination and eclipse depth
and obtain multi-epoch spectroscopic data in order to derive
the fundamental parameters (mass, radius, effective temperature,
metallicity) and distance.

\acknowledgements{

This research has made use of the SIMBAD database, operated at CDS,
Strasbourg, France.

}

{}

\

\newpage

\

\appendix

\section{The model}
\label{AppA}

Compared to the code described in G04 the part of fitting sine and
cosine functions to the lightcurves is omitted as the  shape of the 
lightcurve of an EB is not described by such a functional form.

The determination of the period is done in two steps: using Fourier
analysis (as in G04), and subsequently phase dispersion minimization
(PDM), as described by Stellingwerf (1978), and which is new to the code.

The Fourier analysis is done with the {\sc numerical recipes} (Press
et al. 1992) subroutine {\it fasper}. Inputs to it are the time and
magnitude arrays. In addition one has to specify two parameters, {\it ofac} 
and {\it hifac}, that indicate a ``typical oversampling factor'' and
the maximum frequency in terms of a ``typical'' Nyquist frequency.

In the present work {\it ofac } = 22 (as in G04) and {\it hifac} =
21.0 are used (compared to 0.8 in G04). The latter parameter is the
determining factor in both the computational speed, and the shortest
period that can be found. In this case, it is set to correctly
identify the main period in OGLE sc8/119647 (OGLE005938.46-730002.2)
with a known orbital period of 0.85494 days (Wyithe \& Wilson 2001),
which is the shortest period in the test dataset described in the main text. 
It implies that in the configuration used in the present paper there
is a bias in the detection of orbital periods shorter than about 0.85
days, of no severe consequence as the focus is finding detached EBs.

The outputs of {\it fasper} are the frequency where the peak of the
power spectrum occurs and a number indicating a {\it significance}. 
One of the main parameters in the code is to provide the critical
cut-off above which a period is not considered to be significant. In
the present work {\it significance = $2.2 \times 10^{-5}$ } is used,
and this is determined empirically, by using the set of detached EBs
which the code should be able to retrieve (see below).

As had already been noted in G04 and Groenewegen \& Blommaert (2005)
in analysing LPVs, the frequency returned from the Fourier analysis in
some cases turns out to be a harmonic rather than the true period and
this needs then manual refitting of the lightcurve. The same turns out
to be the case when the test data set of EBs was initially being
analysed.  Therefore it was decided to perform a PDM analysis at
selected frequencies to decrease manual intervention and hence to
allow for a more fully automatic analysis. The $\theta$-statistics as
defined by Stellingwerf (1978) is first calculated at the frequency
$\omega$, being the frequency of the peak of the power spectrum
returned by {\it fasper} divided by 2, as the orbital period should
normally be twice the period returned from the Fourier analysis
because of the primary and secondary eclipse within one orbital
period. Then the $\theta$ quantity is calculated for frequencies
0.4$\omega$, 0.5$\omega$ and $\frac{2}{3}\omega$ and
the frequency resulting in the lowest $\theta$--under the condition that it
should be smaller than 0.93$\theta({\omega})$--
is taken as the true orbital frequency. These numbers have been
derived from analysis of the objects where the Fourier analysis
returned the incorrect orbital period.

\begin{figure*}[b]
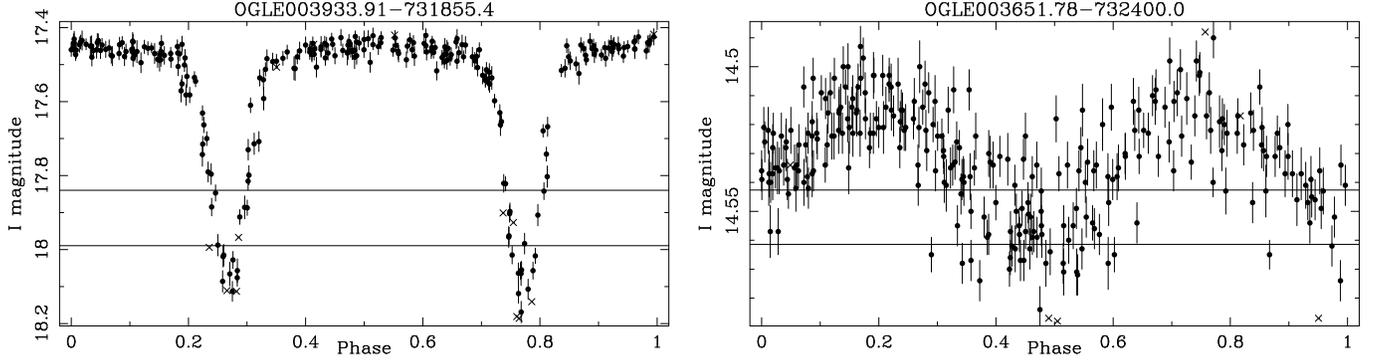


\begin{minipage}{0.49\textwidth}
\resizebox{\hsize}{!}{\includegraphics{EB_example_of_finding.ps}}
\end{minipage}
\hfill
\begin{minipage}{0.49\textwidth}
\resizebox{\hsize}{!}{\includegraphics{EB_example_of_not_finding.ps}}
\end{minipage}
\caption[]{
Example of a phased lightcurve of an EB (left) and an LPV (right), and
the two magnitude limits discussed in the text. The identification of
an EB is based on the number and statistical properties of the
data points fainter than these two limits.
}
\label{Fig-Exa}
\end{figure*}

The lightcurve is phased with the supposed orbital frequency and from
its properties to be described below the candidate EBs are then
selected (see Fig~\ref{Fig-Exa} for an example).

A first limit is set at $m_{\rm lim1} = m_{\rm max} - 0.2\; 
(m_{\rm max} - m_{\rm min})$, where $m_{\rm max}$ and $m_{\rm min}$
represent the faintest and brightest magnitude\footnote{Note that like
in G04 the 3 brightest data points and the 5\% of the data with the
largest photometric error bars are automatically rejected.}. If there
are less than 10 datapoints fainter than $m_{\rm lim1}$ then it is
brightened in steps of 0.04 magnitude until there are.

Then, for a grid of phases $\phi_n$ ($N$ phase steps of 0.01) it is
counted how many of the points fainter than $m_{\rm lim1}$ are within
0.05 in phase of $\phi_n$. The average of this number over the $N$
phase bins defines a ``background'' level {\it bkg}, and the phase
where his number is largest--$n_{\rm max}$--is recorded, $\phi_{\rm max}$.

A second limit is set at $m_{\rm lim2} = m_{\rm lim1} - 0.2$. From
all, $n_2$, phase points fainter than $m_{\rm lim2}$ and within 0.2 in
phase of $\phi_{\rm max}$, a second parameter {\it rms} is determined
as the rms of the phase difference between $\phi_{\rm max}$ and the
phase of the $n_2$ data points. A third parameter {\it rat} is defined
as $n_2 / n_{\rm max}$.

An object is classified as an EB candidate if:
\begin{itemize}
\item $n_2 /$ {\it bkg} $>$ 4.91 and {\it rms} $<0.0373$ and {\it rat} $<2.01$, or

\item $n_2 /$ {\it bkg} $>$ 5.45 and {\it rms} $<0.036$ , or

\item $n_2 /$ {\it bkg} $>$ 5.50 and                        {\it rat} $<1.64$, or

\item                                {\it rms} $<0.022$ and {\it rat} $<2.34$.

\end{itemize}

The principal idea behind this procedure is that compared to the
phased light curve of a pulsational variable the one of an EB has at
least one minimum, and that the spread in phase at this minimum is
limited, and that this also holds at the second magnitude limit
(because of the limited width of the eclipse).

The exact value of the parameters have been derived from analysis of a
test data set, as described in the main text, and with that setting
the program correctly selected 137 out of a 141 objects (= 97\%).

\

\newpage

\

\section{Newly discovered EBs in the MCs}
\label{AppB}

\begin{figure*}
\includegraphics[width=175mm]{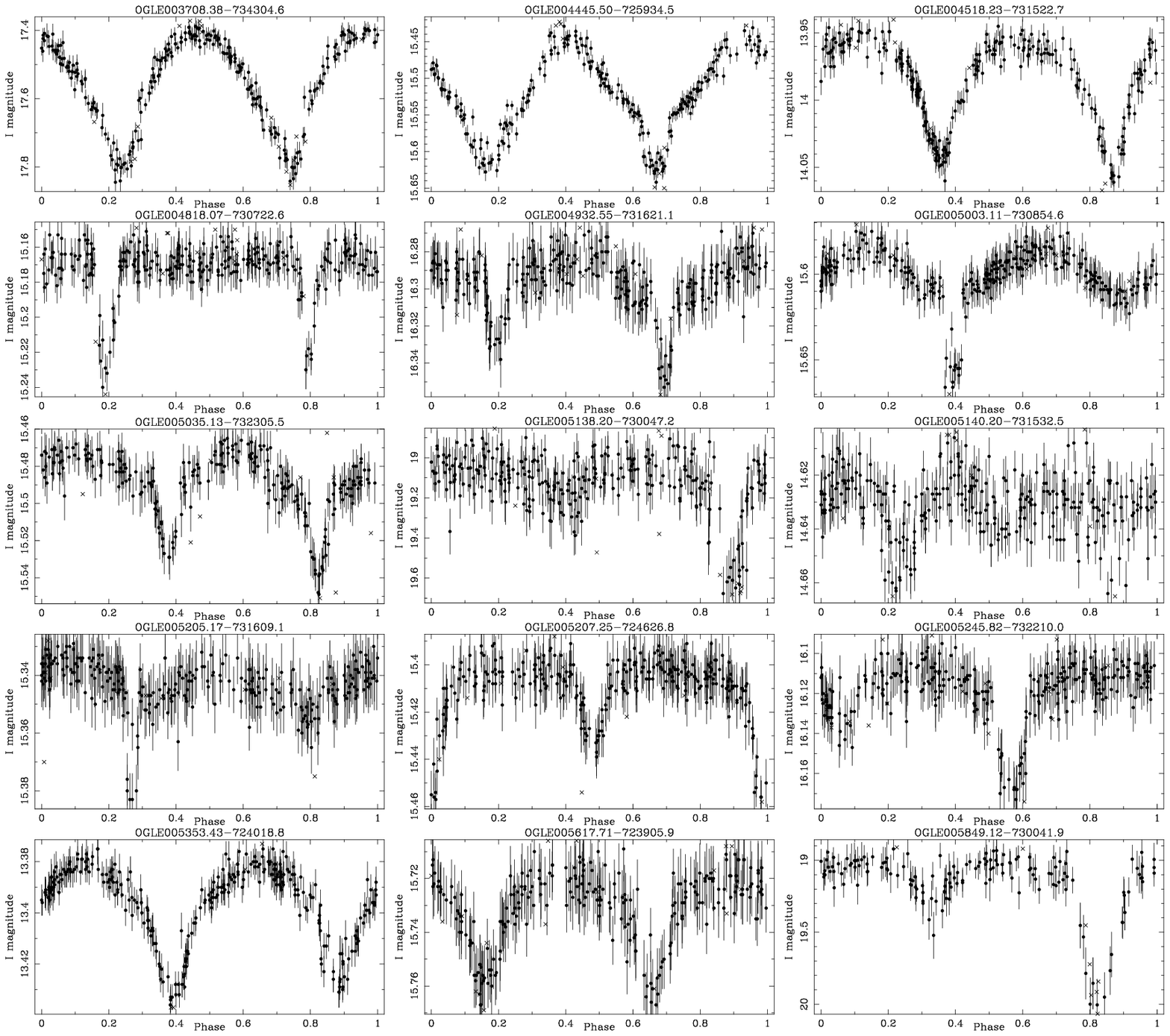}
\caption[]{
First entries of electronically available figure with the phased
lightcurves of newly discovered EBs in the MCs. Phase zero
is given by the Julian Date listed in Table~\ref{TAB-NewMCs}.
}
\label{Fig-EB-NEW}
\end{figure*}

This appendix lists the names and periods (Table~\ref{TAB-NewMCs}) and
phased lightcureves (Figure~\ref{Fig-EB-NEW}) of the newly identified
EBs in the SMC and LMC.


\begin{table*}
\caption{Properties of newly discovered EBs in the MCs.}
\begin{tabular}{lcrr} \hline
   Field  &          OGLE name    & Period ($d$) & $T^{(a)}$ \\ \hline

 smc\_sc1 & OGLE003708.38-734304.6 &   8.720979 & 627.870 \\ 
 smc\_sc3 & OGLE004445.50-725934.5 & 204.406036 & 625.927 \\ 
 smc\_sc3 & OGLE004518.23-731522.7 &   1.002808 & 621.836 \\ 
 smc\_sc4 & OGLE004818.07-730722.6 &   3.867470 & 622.874 \\ 
 smc\_sc5 & OGLE004932.55-731621.1 &   2.611493 & 466.548 \\ 
 smc\_sc5 & OGLE005003.11-730854.6 & 197.654480 & 466.548 \\ 
 smc\_sc5 & OGLE005035.13-732305.5 &   3.332872 & 466.548 \\ 
 smc\_sc6 & OGLE005138.20-730047.2 &   6.514974 & 466.571 \\ 
 smc\_sc6 & OGLE005140.20-731532.5 &  91.864151 & 466.571 \\ 
 smc\_sc6 & OGLE005205.17-731609.1 & 256.679626 & 466.571 \\ 
 smc\_sc6 & OGLE005207.25-724626.8 &   1.763637 & 466.571 \\ 
 smc\_sc6 & OGLE005245.82-732210.0 &   2.961430 & 466.571 \\ 
 smc\_sc6 & OGLE005353.43-724018.8 &   2.950854 & 466.571 \\ 
 smc\_sc7 & OGLE005617.71-723905.9 &   1.554488 & 626.927 \\ 
 smc\_sc8 & OGLE005849.12-730041.9 &   1.424265 & 626.935 \\ 
smc\_sc10 & OGLE010521.71-723740.0 & 136.450470 & 621.867 \\ 
lmc\_sc15 & OGLE050100.28-692243.8 &   1.125093 & 726.815 \\ 
lmc\_sc15 & OGLE050100.36-690534.7 &  27.095217 & 726.815 \\ 
lmc\_sc15 & OGLE050124.20-683728.9 &   4.690511 & 726.815 \\ 
lmc\_sc15 & OGLE050124.33-690033.6 &   5.216741 & 726.815 \\ 
lmc\_sc15 & OGLE050129.81-683647.0 &   5.018933 & 726.815 \\ 
lmc\_sc14 & OGLE050301.95-685526.9 &   9.183692 & 726.806 \\ 
lmc\_sc14 & OGLE050337.43-685622.7 &   1.512742 & 726.806 \\ 
lmc\_sc14 & OGLE050415.35-692123.5 &   3.679661 & 726.806 \\ 
lmc\_sc14 & OGLE050428.68-685525.7 &  29.598274 & 726.806 \\ 
lmc\_sc14 & OGLE050438.66-685410.7 &   2.435237 & 726.806 \\ 
lmc\_sc14 & OGLE050440.04-692235.7 &   1.161064 & 726.806 \\ 
lmc\_sc13 & OGLE050505.98-690310.0 &   1.048826 & 726.798 \\ 
lmc\_sc13 & OGLE050511.85-685947.9 &   3.438104 & 726.798 \\ 

\hline
\end{tabular}

(a) $T$ = (JD-2450000) used to define zero phase in Figure~\ref{Fig-EB-NEW}.

First entries only. Complete table available in electronic form at the CDS.

\label{TAB-NewMCs}
\end{table*}


\end{document}